# Networked Sensing for Radiation Detection, Localization, and Tracking

R.J. Cooper[1*], N. Abgrall[1], G. Aversano[1], M.S. Bandstra[1], D. Hellfeld[1], T.H. Joshi[1], V. Negut[1], B.J. Quiter[1], E. Rofors[1], M. Salathe[1], K. Vetter[1,2], P. Beckman[4], C. Catlett[4], N. Ferrier[4], Y. Kim[3], R. Sankaran[3], S. Shahkarami[3], S. Amitkumar[4], E. Ayton[4], J. Kim[4], S. Volkova[4]

[1]Lawrence Berkeley National Laboratory, USA, [2]The University of California, Berkeley, USA, [3]Argonne National Laboratory, USA, [4]Pacific Northwest National Laboratory, USA

*rjcooper@lbl.gov

**Abstract**. The detection, identification, and localization of illicit radiological and nuclear material continue to be key components of nuclear non-proliferation and nuclear security efforts around the world. Networks of radiation detectors deployed at strategic locations in urban environments have the potential to provide continuous radiological/nuclear (R/N) surveillance and provide high probabilities of intercepting threat sources. The integration of contextual information from sensors such as video, Lidar, and meteorological sensors can provide significantly enhanced situational awareness, and improved detection and localization performance through the fusion of the radiological and contextual data. In this work, we present details of our work to establish a city-scale multi-sensor network testbed for intelligent, adaptive R/N detection in urban environments, and develop new techniques that enable city-scale source detection, localization, and tracking.

## 1. Introduction

The ability to detect, identify, localize, and track radiological and/or nuclear (R/N) material which is out of regulatory control and is present in, or transiting through, an urban environment are key components of nuclear security efforts around the world. Technical challenges include the detection of weak and/or shielded sources during short dwell-time encounters, the presence of large, spatially and temporally varying backgrounds, and the need for very low false positive rates (e.g. one per 8 hours or lower) [1]. Practical challenges include the significant population of industrial and medical "nuisance" sources which are typically present in urban environments, and which must be adjudicated when detected, and the need to cover large areas with finite resources. The process of monitoring an urban environment for the presence of illicit R/N material is often referred to as "R/N wide area search" and typically includes the use of mobile and static gamma-ray and neutron detector systems.

The concept of employing distributed networks of detectors has long been discussed, and explored theoretically (e.g., Ref. [2]), but never previously realized on a meaningful scale. Important examples include the DARPA SIGMA project [3] which demonstrated the deployment of large numbers of static and mobile detectors and the use of cloud computing to process the data in near real-time, and the NOVArray project [4] in which an array of eighteen large volume gamma-ray detectors were placed at traffic intersections in Northern Virginia in order to explore radiological background dynamics and nuisance sources observations. One of the ways in which these systems both differ from true networks is that each detector operates independently. In a true sensor network, each sensor is connected such that information is able to flow between them (possibly via a central node), allowing optimization to be performed at the level of both individual sensors and the system as a whole.

In this work, we describe ongoing efforts to establish a multi-sensor testbed for the development and evaluation of advanced concepts in intelligent, networked radiation detection. This network combines gamma-ray detectors with contextual sensors including video, Lidar, and environmental sensors, real time edge computing at each sensor node, inter-node connectivity, and network-level computation in the cloud. In Section 2 we lay out our vision for intelligent networked radiation detection. In Section 3 we describe the sensor network that is being developed and fielded in Chicago, and provide details of the current status of some of the key algorithms. Finally, in Section 4, we discuss the future directions for this project and this area of research in general.

## 2. A Vision for Networked Radiation Detection

Networked radiation detection systems have the potential to offer wide area coverage and continuous radiological surveillance, afford improved detection sensitivity relative to individual detectors or the type of non-connected arrays which have been recently demonstrated, allow an increased ability to interpret and adjudicate observed anomalies, and provide increased radiological domain awareness in complex environments such as cities. By placing static detectors at strategic locations such as natural chokepoints or major intersections, a high probability of encountering mobile sources can be achieved. An advanced detector network should also have the ability to adapt to changing conditions, fuse data from multiple sensors in the network, and leverage spatiotemporal correlations for improved performance. To this end, we are developing a testbed of multi-sensor systems with which to explore these concepts. The key characteristics of this network are:

- High-sensitivity, *real-time R/N anomaly detection* and isotope identification.
- Integrated *contextual sensing* to provide increased situational awareness, improved ability to interpret radiological anomalies, and improved source detection, localization, and tracking.
- AI at the edge to enable *dynamic adaptation of sensors and algorithms*.
- *Network-level fusion* of radiological and contextual data to enable improved sensitivity to weak sources and increased domain awareness.
- *Inter-node communication* to enable system-level adaptation and optimization.
- *Spatiotemporal correlation* of radiological, contextual, and pattern-of-life information.
- Hardware and software approaches that are applicable *from street-scale to city-scale* networks.

## 3. The PANDAWN Network

The sensor network is being developed and fielded in the context of two collaborating projects; the Platforms and Algorithms for Networked Detection and Analysis (PANDA) project which is led by Lawrence Berkeley National Laboratory (LBNL), and the Domain Aware Waggle Network (DAWN) project led by Argonne National Laboratory (ANL). The PANDAWN network will ultimately comprise up to twenty multi-sensor nodes mounted on utility poles throughout the city of Chicago. The PANDAWN nodes are built on the Waggle platform [5], a state-of-the art programmable edge computing and sensing-actuation platform for science. Two sensor systems have so far been deployed in Chicago, with a further sixteen to eighteen planned for deployment by Spring 2023.

3.1. Sensor Systems

Figure 1 shows an engineering drawing of a PANDAWN sensor system (left) along with an image of one of the first two systems deployed on a traffic signal pole in Chicago (right). Each sensor system features a 2x4x16" NaI(Tl) detector which is read out in list-mode. The detector is contained within a custom-designed enclosure which provides protection from the elements and includes heating and cooling mechanisms which maintain an operating temperature between 0° C and 30° C. The detector housing is affixed to the pole via a variable angle mount. The contextual sensor suite includes a 5 MP video camera which operates at 30 frames per second, and a 64-beam Lidar with a 90° vertical field-of-view. The Lidar is co-located with the camera via a variable-tilt mount which allows the sensor to be positioned to maximize the beam coverage on the ground in different deployment locations. In addition to the video and Lidar, there is a microphone, a suite of meteorological sensors (i.e. temperature, pressure, humidity), and a dedicated rainfall sensor. At the heart of the system is a dedicated edge-computing platform that includes two Nvidia Xavier NX edge computers. This unit distributes power to all sensors, allows the data from each sensor to be processed in real-time, and enables communication to the cloud via ethernet, Wi-Fi, or cellular connection. A custom software framework allows the deployment and management of algorithm modules, referred to as *plugins*, which are implemented as Docker containers.

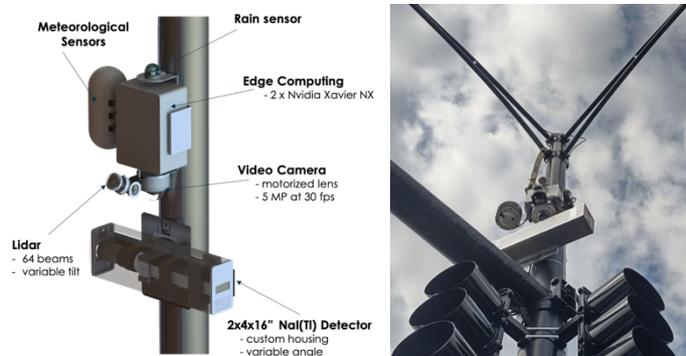

Figure 1. An engineering drawing of a PANDAWN multi-sensor system (left), along with a photograph of one of the first two systems deployed in Chicago (right).

3.2. Node-level Algorithms

Multiple algorithms to process the data streams from the various sensors have been developed and implemented. This development has taken place using data from the NOVArray and purpose-built systems at LBNL. As these algorithms will be implemented on each sensor system, they are referred to as *node-level algorithms*. These algorithms are intended to provide real-time source detection, identification, and localization, provide a high degree of situational awareness, and also provide information which can inform network-level data fusion and decision making. Specific algorithms include a continuous, real-time energy calibration algorithm, radiological anomaly detection and isotope identification based on the use of Non-negative Matrix Factorization (NMF) [6], and a framework for object detection and tracking using both video and Lidar, and the attribution of anomalous radiological signals to tracked objects.

The energy calibration algorithm is intended to maintain a stable energy calibration in the face of temperature-induced changes in the detector response. The approach continuously fits the recorded background spectra to a series of templates that describe the detector response to the various background terms (i.e., K-40, U and Th series, cosmic rays, etc.), while also fitting a multi-parameter calibration model. This approach has been shown to reduce the variability in the energy calibration to less than 0.5% in the temperature range -20° C to 40° C. The real-time anomaly detection and isotope identification algorithm is based on the NMF-based method presented in Ref. [7], which has previously been shown to provide high detection sensitivity at low false alarm rates. The specific implementation developed for PANDAWN includes a number of modifications that allow the detection algorithms to automatically adapt to changing conditions or new deployment environments. These modifications include the ability to build the NMF model of the background, upon which the algorithm relies, *ab initio* and to automatically update if and when the background environment changes, and the ability for the isotope specific detection thresholds to be modified either automaticall in light of the changing background or according to instructions received from elsewhere in the network, and delivered via the cloud. Ongoing developments include methods to update the library of isotope templates used for isotope identification based on real-world data. Details of these approaches can be found in Ref. [8].

The primary contextual sensor processing pipeline is based on the object detection, tracking, and source-object attribution process described in Ref. [9]. Ongoing developments include updating the machine learning algorithms that perform object detection, improving the object tracking algorithm, and exploring methods to extract additional information from the video data e.g., vehicle make, model, and color.

3.3. Cloud Algorithms

Data from a given PANDAWN node is transferred to the cloud either at a fixed frequency or based on a specific triggering event (e.g., a radiological anomaly). At the current time, the complete data streams from all sensors are stored but it is envisioned that in the future, significant data reduction will occur at the edge. Executing certain algorithms in the cloud allows a) computationally intensive processes and/or processes that do not need to be performed in real time to be shifted away from the edge and b) the implementation of algorithms that act on data from multiple sensor nodes. The ability to perform node-triggered re-training of machine learning models (e.g., NMF background models, or models for object detection) is currently under development and will be in the field soon. Cloud algorithms which are currently being explored include radiological and contextual data fusion including object redetection, network-level detection, and probabilistic source tracking. Data from the NOVArray and the SIGMA program have also been used to develop predictive analytics methods that employ machine learning methods to predict the rate of observation of medical and industrial nuisance sources using historical priors and open source data such as construction permits, hospital locations, and traffic data [10]. These methods have the potential to increase domain awareness, provide important context to support the interpretation and adjudication of ambiguous radiological anomalies, and could potentially also be used to inform isotope identification algorithms.

3.4. Towards Network-Level Detection and Tracking

Two of the most compelling attributes of a detector network are the potential to provide improved detection sensitivity through the fusion of data from multiple nodes, and the ability to track R/N sources as they move through the network. Both of these concepts are currently being explored using a combination of experimental data and highly detailed simulations. Our initial analyses suggest that the fusion of radiological data from multiple observations of a given source (i.e. network-level detection) can significantly improve the sensitivity to weak sources, relative to performing detection on a per-detector basis (as in a more conventional distributed array). However, in order to realize this gain in detection performance, the network must leverage detailed contextual information in order to track the source and identify the time windows over which to aggregate the data from multiple sensor nodes. In the case of a source in a vehicle, this includes information such as the make, model, and/or color of the vehicle, which can be extracted from the real-time video recorded at each sensor node. The image in Figure 2 shows results from agent-based simulations in which vehicle-borne, 5 µCi Cs-137 sources were in transit through a 6 km² area of central Chicago in the presence of PANDAWN-like detector networks of different size (i.e., different numbers of detectors). The result shows the probability of detection as a function of network size, as different types of contextual information are used to identify correlated observations, with these correlated observations being subsequently fused to perform network-level detection. The performance of a more conventional detector array, where data fusion does not take place and the overall detection performance is defined by the performance of the individual detectors, is shown by the dashed brown line. These results demonstrate that the detection performance of the network is significantly improved when information such as the vehicle color, model, or the combined make and color are exploited.

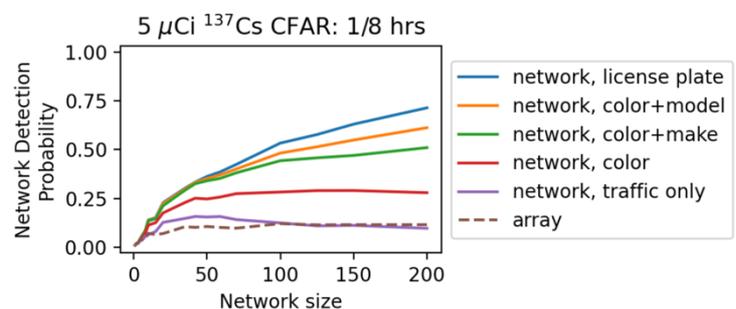

Figure 2. Results of an agent-based simulation in which vehicle-borne, 5 µCi Cs-137 sources were in transit through Chicago in the presence of PANDAWN-like detector networks of different size.

They also show that using traffic information alone does not provide sufficiently precise source tracking to enable robust data fusion. In some cases, the use of traffic data to inform radiological data fusion results in worse performance than operating the network as a simple array. Fundamentally, these modelling studies suggest that networked detection has the potential to significantly improve detection performance relative to non-networked systems, but using intelligent algorithms and leveraging contextual data is essential if this is to be realized.

## 4. Conclusions and Future Directions

We are developing and deploying a novel, multi-sensor network that combines state-of-the-art sensing, advanced algorithms, and edge and cloud computing to serve as a testbed for the development and demonstration of new concepts in networked radiation detection. Through the integration of contextual information, adaptive algorithms, and data fusion methodologies, the research performed with this platform will evaluate the feasibility of networked detection systems to provide new capabilities for R/N detection in urban environments. The PANDAWN network will scale from two systems to eighteen or twenty by Spring 2023 and will include some minor changes to the hardware. The node level algorithm suite will be deployed on all systems and evaluated using the acquired data. Two-way communication between the nodes, and cloud-driven adaptation of node-level algorithms and sensors will be demonstrated in mid 2023. Key areas for future exploration include utilizing new machine learning methods to analyse and fuse multi-modal data and exploit temporal and spatial patterns for improved detection and localization, exploring how historical data trends can be used to inform and enhance the detection process, and establishing the optimum balance between edge and cloud processing.

**Acknowledgements**
This work was performed under the auspices of the U.S. Department of Energy by Lawrence Berkeley National Laboratory (LBNL) under Contract DE-AC02-05CH11231. The project was funded by the U.S. Department of Energy, National Nuclear Security Administration, Office of Defense Nuclear Nonproliferation Research and Development.